# Comparative Analysis of Lightweight Kubernetes Distributions for Edge Computing: Performance and Resource Efficiency

Diyaz Yakubov[0009-0000-7989-4614] and David Hästbacka[0000-0001-8442-1248]

Tampere University, Tampere FI-33014, Finland
{diyaz.yakubov, david.hastbacka}@tuni.fi

**Abstract.** Edge computing environments increasingly rely on lightweight container orchestration platforms to manage resource-constrained devices. This paper provides an empirical analysis of five lightweight kubernetes distributions (KD) — k0s, k3s, KubeEdge, OpenYurt, and Kubernetes (k8s) — focusing on their performance and resource efficiency in edge computing scenarios. We evaluated key metrics such as CPU, memory, disk usage, throughput, and latency under varying workloads, utilizing a testbed of Intel NUCs and Raspberry Pi devices. Our results demonstrate significant differences in performance: k3s exhibited the lowest resource consumption, while k0s and k8s excelled in data plane throughput and latency. Under heavy stress scenarios, k3s and k0s accomplished the same workloads faster than the other distributions. OpenYurt offered balanced performance, suitable for hybrid cloud-edge use cases, but was less efficient in terms of resource usage and scalability compared to k0s, k3s and k8s. KubeEdge, although feature-rich for edge environments, exhibited higher resource consumption and lower scalability. These findings offer valuable insights for developers and operators selecting appropriate KD based on specific performance and resource efficiency requirements for edge computing environments.

## 1 Introduction

In recent years, the edge computing demand has grown significantly resulting in more research [11] [10] [7] and springing up new software solutions to cope with new issues. The enablers of such growth are improvements in communication (5G, such as 5G IoT [2]) and ways of managing/maintaining the software on the edge devices (containerization). While many data systems successfully pioneered the orchestration techniques for containerization, the Internet of Things (IoT) was lagging in that area, though many of the kubernetes distributions (KD) can be used for it. However, IoT has slightly different requirements that are not widely considered in common-purpose orchestration tools, such as Kubernetes (k8s). To benefit from battle-tested orchestration of k8s technology and to address IoT-specific requirements on resource-constrained devices, many lightweight KDs have been devised recently.

Lightweight KDs like k0s, k3s, KubeEdge, and OpenYurt are evolving rapidly, each with unique features and components. This diversity complicates the choice for developers, especially for resource-constrained edge devices in factories, autonomous vehicles, or smart cities, where container orchestration overhead can be critical. Application performance on these distributions is influenced by factors like container runtime, control plane, data plane storage, and networking. These factors are crucial for assessing resource efficiency, analyzing cluster behavior under stress, and optimizing resource usage.

Performance and capacity planning for microservice architectures are key challenges in the performance engineering community. Previous studies [3, 8, 9] have compared full and managed KDs, but research on lightweight distributions like k0s and k3s is limited and often inconsistent. Newer distributions like k0s and OpenYurt remain understudied.

This paper presents an empirical study of popular lightweight KDs. We evaluated resource utilization, throughput and response times under stress scenarios. Our benchmarking setup involved two Intel NUCs (Next Unit of Computing, small form factor computer) and three Raspberry Pi 4 Model B single-board computers, netdata for data collection, MongoDB for storage, Python for analysis and visualization, and Ansible with bash scripts for automation and provisioning which can be replicated by other researchers. The following research questions (RQ) were formulated to explain the test results and provide a comprehensive analysis:

- RQ1 Resource Utilization Comparison: Assess and compare the resource efficiency of various lightweight KDs when deployed on resource-constrained devices, such as Raspberry Pis. Focus on metrics like CPU, memory, and storage utilization to determine which distributions are most suitable for such environments.
- RQ2 Cluster Behavior Characterization: Analyze the behavior of kubernetes clusters under various stress conditions, including: 1) Light and heavy CPU loads to gauge processing efficiency. 2) Network-intensive activities to understand bandwidth and latency implications.
- RQ3 Resource Optimization: Identify which KD is most suitable for specific environments, focusing on efficient workload scheduling, auto-scaling, and resource management to optimize performance on constrained devices.

The results will provide valuable guidance for both practitioners and researchers selecting KDs in edge computing environments.

The paper is structured as follows: Section 2 describes KDs and reviews previous performance analyses. Section 3 details the experimental setup. Section 4 presents the benchmarking results answering research questions, followed by Section 5 which discusses the results. Section 6 concludes the paper and highlights future work.

## 2 Background and Related Work

### 2.1 Kubernetes distributions under test

Kubernetes has become the de facto standard for container orchestration, automating deployment, scaling, and management of containerized applications. Various distributions have been developed for different use cases, particularly resource-constrained environments like edge devices and IoT gateways. This section overviews notable distributions: k0s, k3s, k8s, OpenYurt, and KubeEdge, each offering varying levels of complexity, resource consumption, and features tailored to specific deployment needs, from robust cloud infrastructures to resource-limited edge devices.

Kubernetes[1] is an open-source orchestration platform originally developed by Google and now maintained by the Cloud Native Computing Foundation (CNCF). It is designed to automate deploying, scaling, and operating application containers supporting a broad ecosystem of tools for monitoring, logging, networking, security, etc.. Still, its resource intensity makes it less suitable for limited-resource environments.

k3s[2], developed by Rancher Labs (now part of SUSE), is a lightweight KD designed for edge computing, IoT, and CI/CD pipelines. Packaged as a single binary (around 100 MB), it simplifies the installation process and requires less memory and CPU. Likewise, it supports ARM processors making it suitable for devices like Raspberry Pi.

k0s[3], created by Mirantis, is another lightweight distribution focused on minimal resource consumption and ease of installation. It supports various storage options like etcd and SQLite, and it is designed for bare metal, edge, and cloud environments, emphasizing security and versatility across ARM and x86 platforms into a single binary file.

OpenYurt[4], developed by Alibaba Cloud, extends Kubernetes to edge computing, enhancing it for cloud-edge hybrid environments. It supports edge autonomy and edge-cloud synergy with features like YurtHub for traffic routing and YurtTunnel for secure communication between cloud and edge nodes, making it ideal for smart cities, industrial IoT, and remote monitoring.

KubeEdge[5], a CNCF project, extends Kubernetes to edge computing environments, offering infrastructure and APIs to manage applications on edge nodes. It supports offline autonomy, allowing edge nodes to function independently of the cloud, and simplifies IoT and Industrial Internet communications with components like EdgeHub for device communication and EdgeController for managing edge nodes from the cloud.

---

1 https://kubernetes.io/
2 https://k3s.io/
3 https://k0sproject.io/
4 https://openyurt.io/
5 https://kubeedge.io/

### 2.2 Related work

Over the last years, performance studies have been conducted on lightweight KDs. Each cover specific KDs and metrics thoroughly, though there are still areas that require further investigation. Following, we describe previous works and discuss the differences with our work.

Koziolek et al. [6] compared lightweight KDs, specifically Microk8s, k3s, k0s, and MicroShift, focusing on resource usage as well as control plane and data plane performance under stress scenarios. They found that k3s and k0s had slightly higher control plane throughput, while MicroShift excelled in data plane throughput, providing useful insights for selecting distributions.

Cilic et al. [12] evaluated Kubernetes, k3s, KubeEdge, and ioFog in edge computing, assessing deployment complexity, memory footprint, and performance. Kubernetes stood out with its low memory footprint and strong performance, but the study also noted specific challenges for each tool in edge environments.

Kjorveziroski and Filiposka et al. [5] examined KDs for serverless edge computing using OpenFaaS, finding k3s and Microk8s performed best in most benchmarks, while full Kubernetes excelled under sustained loads.

Fogli et al. [4] assessed KDs in tactical networks with limited bandwidth and high latency, concluding that KubeEdge outperformed k8s and k3s in maintaining cluster stability under degraded conditions, making it ideal for tactical applications.

Bahy et al. [1] compared KubeEdge, k3s, and Nomad, focusing on resource utilization in edge computing. They found Nomad was the most efficient in CPU and memory usage, while k3s excelled in storage efficiency, offering insights into choosing the best container orchestrator for resource-constrained environments.

While the provided works offer valuable insights into the performance of various lightweight KDs, our research differentiates itself by conducting detailed stress tests covering light and heavy CPU loads, and network-intensive activities on real devices with some KDs that are not well covered, such as k0s, KubeEdge and OpenYurt. By addressing these aspects, our work aims to provide a more holistic understanding of the performance and efficiency of lightweight KDs in edge computing environments.

## 3 Methodology

### 3.1 Test setup and equipment

In this section, we detail the equipment and configurations utilized to empirically evaluate several lightweight KDs, including Kubernetes itself: k0s, k3s, k8s, OpenYurt, and KubeEdge, listed in Table 1 [6] [7] [8].

[6] KubeEdge and OpenYurt extend the k8s to the edge, consequently, the cloud part is still requiring CNI to operate and dictate what control plane storage is.

[7] https://github.com/kubeedge/edgemesh

[8] https://openyurt.io/docs/core-concepts/raven/

**Table 1.** Setup attributes of kubernetes distributions

| Feature | k0s | k3s | Kubernetes (k8s) | KubeEdge | OpenYurt |
|---|---|---|---|---|---|
| Version | v1.28.4+k0s.0 | v1.27.4+k3s1 | v1.29 | 1.14.4 | v1.4.0 |
| Container Network Interface (CNI) | kube-router (default, v1.1.1) | flannel (default, v0.22.0) | flannel (v0.24.3) | flannel[7](v0.24.3) + edgemesh[8] (v1.14.0) | flannel[7](v0.24.3) + raven[9] (0.4.1) |
| Container Runtime Interface (CRI) | containerd (v1.7.11) | containerd (v1.7.11) | containerd (v1.7.11) | containerd (v1.7.11) | containerd (v1.7.11) |
| Control Plane Storage | etcd(v3.5.10) | SQLite(3.39.2) | etcd(v3.5.10) | etcd[7](v3.5.10) | etcd[7](v3.5.10) |
| Type of distribution | lightweight | lightweight | full-fledged | edge extension for k8s | edge extension for k8s |

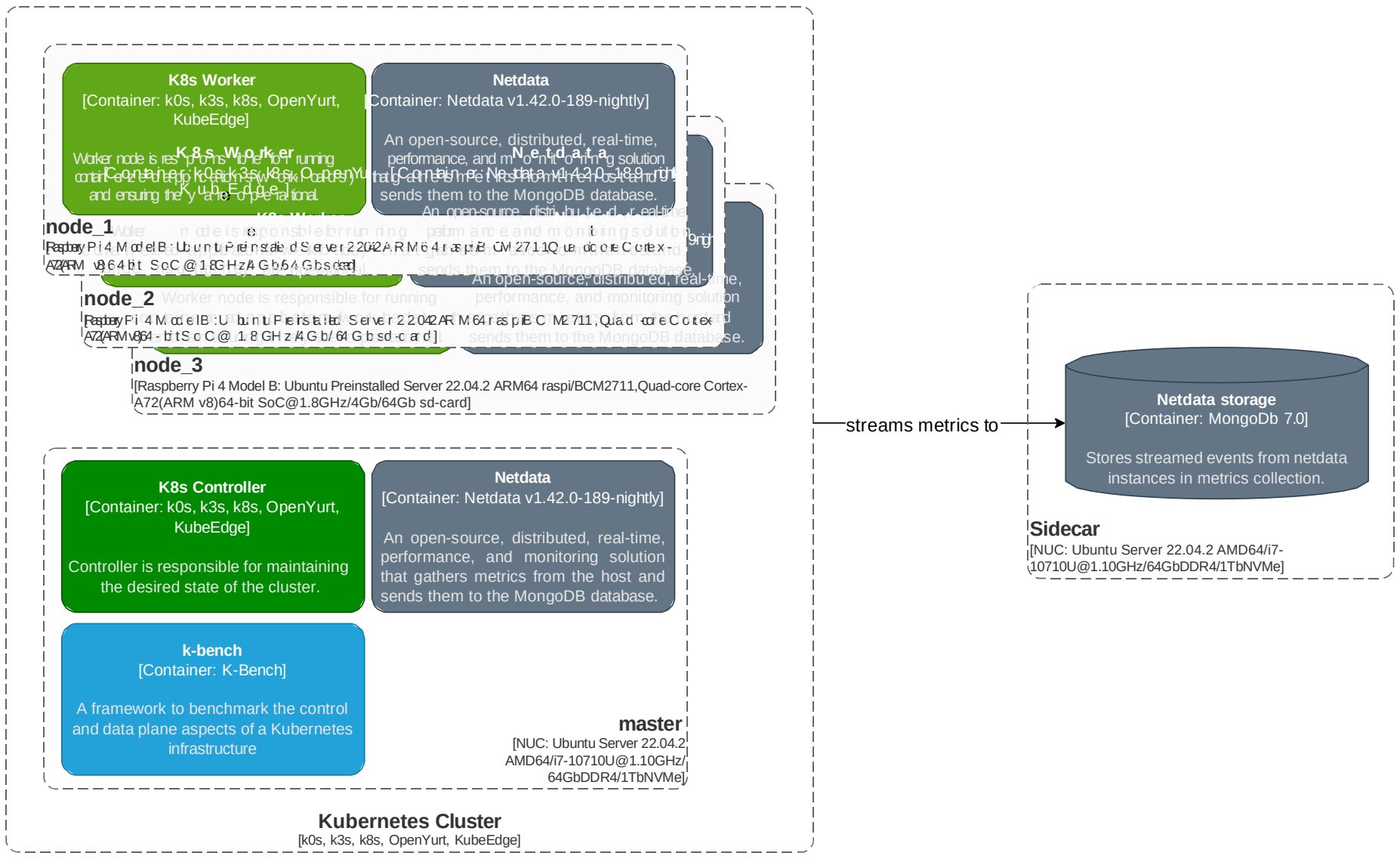

**Fig. 1.** Deployment diagram

Our primary objective is to gauge and compare resource utilization, specifically focusing on CPU, network bandwidth, disk I/O, and RAM. To provide a holistic view, we incorporate test cases that load kubernetes clusters differently in order to overview the consumption of resources and the cluster behavior in different use cases. Figure 1 depicts the setup deployment view.

In the diagram, there are several NUC and Raspberry Pi 4 Model B (RPi) machines. One NUC (master) and three RPi (workers) machines form the kubernetes cluster, while the second NUC machine acts as a side container machine monitoring and storing data. Apart from KDs, all machines in this cluster have netdata installed to gather metrics and stream them to the sidecar machine to the MongoDB database. Likewise, the master machine has additional software installed for performing performance tests (k-bench [9]).

9 https://github.com/vmware-tanzu/k-bench with the last commit's sha hash: 53a82d316effaaf562d81a7cd306bf5f0d40cfc6

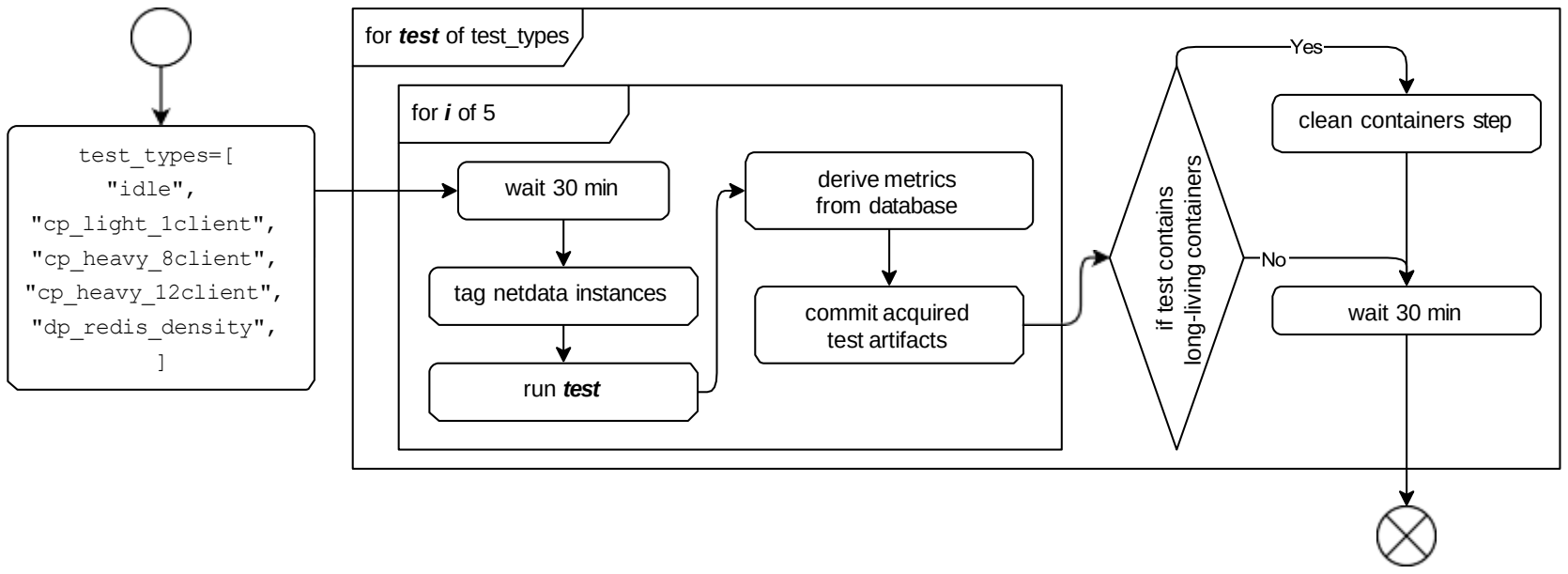

**Fig. 2.** Test flow diagram

### 3.2 Test procedures

The Activity Flow diagram in Figure 2 illustrates the testing process. Each test type is executed 5 times, with each cycle starting with a 30-minute wait to stabilize the system. The execution tags all Netdata streaming instances for data validation, followed by the test execution using Ansible scripts combined with bash. Data is fetched from the sidecar machine's MongoDB database via an Ansible script, and the results are committed to the git repository to preserve information and signal test completion. Two post-test steps follow: cleaning up containers (for tests with long-running containers) and another 30-minute wait before the next test cycle. Test suites are fully automated and can be run per KD. Table 2 lists all test scenarios, their descriptions, used tools, and collecting metrics.

**Table 2.** Test scenarios to benchmark performance of a cluster

| **Test Name - Description** | **Tools** | **Collecting Metrics** |
|---|---|---|
| **idle** - Gauges how much resources the cluster consumes when there is no workload, in its serenity state within 5 min. | bash, ansible | CPU, RAM, Network bandwidth, disk I/O |
| **cp_light_1client** - Executes lifecycle actions (CREATE, LIST, GET, UPDATE, DELETE) on Pods, Deployments, Namespaces, and Services to put a light load. Includes specific sleep intervals post-CREATE for timing analysis, with no cleanup post-test. Deployments feature 5 replicas using k8s.gcr.io/pause:3.1 image; all operations are executed once per resource type. | bash, ansible, k-bench | CPU, RAM, Network bandwidth, disk I/O |
| **cp_heavy_8client** - Conducts extensive operations on Pods, Deployments, Namespaces, and Services, executing 8 cycles of CREATE, LIST, GET, UPDATE, DELETE actions to put a considerable load. Pods and Deployments have initial sleep times of 20 and 40 seconds post-CREATE, respectively, to assess timing dynamics, without cleanup post-execution. Deployments are configured with 5 replicas using k8s.gcr.io/pause:3.1, and all resources undergo the same set of actions 8 times. | bash, ansible, k-bench | CPU, RAM, Network bandwidth, disk I/O |
| **cp_heavy_12client** - Performs 12 cycles of CREATE, LIST, GET, UPDATE, DELETE operations on kubernetes Pods, Deployments, Namespaces, and Services to put a stress load. Pods start with a 30-second pause post-CREATE, Deployments with 60 seconds, aimed at deeper timing analysis, without post-test cleanup. Each Deployment configures 5 replicas using k8s.gcr.io/pause:3.1. The test iterates through each action 12 times for each resource, designed for rigorous performance assessment. | bash, ansible, k-bench | CPU, RAM, Network bandwidth, disk I/O |
| **dp_redis_density** - Executes a series of Pod operations within a 1-minute timeframe, focusing on a Redis workload across 3 cycles without cleanup to put a data-heavy load. Initially, 3 Pods with the nginx image and Redis-specific configurations are created, each followed by a 100-second pause. A precondition checks for a specific file in a pod before running Redis server commands in matching Pods, with a 5-second pause post-execution. Another set of operations runs benchmarking commands in Redis worker Pods, also followed by a 5-second pause. Finally, outputs from the Pods are copied locally after a 20-second wait. | bash, ansible, k-bench | CPU, RAM, Network bandwidth, disk I/O, Throughput operations, Latency |

### 3.3 Limitations

While our methodology is comprehensive, certain limitations are acknowledged: *Hardware Constraints*: The use of specific devices (Intel NUCs and Raspberry Pis) may influence performance results; however, they are representative of common edge computing hardware although limited in number. Additionally, homogeneous worker nodes allow for minimizing variables and focus on the KDs themselves. *Network Conditions*: Tests were conducted in a controlled network environment, which may not fully capture the variability of real-world edge networks.

## 4 Results

While k0s, k3s, and k8s solve common issues of orchestration and distribution, KubeEdge and OpenYurt precisely extend them to the edge devices by providing more IoT-grained features, such as on-device storage, digital twin, etc. Of course, those extra features add overhead to the distributions that might slow down their performances.

### 4.1 Light tests

The diagrams in Figure 3 provide focus on the idle and cp_light_1client tests for testing distributions. It is worth noting that in these two tests, the execution times of systems were considered as well, nevertheless, the time results remain the same across distributions for idle and cp_light_1client tests, exactly 5 min and 4 min respectively. Therefore, a timeline was excluded from the analysis for the light test cases.

It is visible that on a master node, the resource consumption metrics follow the same pattern despite the chosen test case showing a slight increase by approximately 0.3% or 0.5% across resources for the cp_light_1client test. The behavior of worker nodes presents different resource consumption patterns, specifically, the CPU and Disk IO usages increase significantly for all distributions under the cp_light_1client test workload, while the RAM and Network utilizations remain the same. In fact, the increased metrics show double gains in consumption. Thus, adding a lightweight task to an idle cluster shows that KDs scale their resource consumption differently. For instance, the CPU of k0s, k3s, k8s, and OpenYurt scale up similarly, adding roughly 2% of CPU consumption while KubeEdge adds almost 4%. The Disk usage also resembles the growing pattern for k0s, k3s, k8s, and OpenYurt distributions by representing around 3% of growth when the KubeEdge rockets up by 6% reaching 9% of Disk utilization. Besides that, the only distribution that continuously uses the network is k0s. It can be justified by Konnectivity[10] service that is exploited in k0s for the Controller to Worker communication[11].

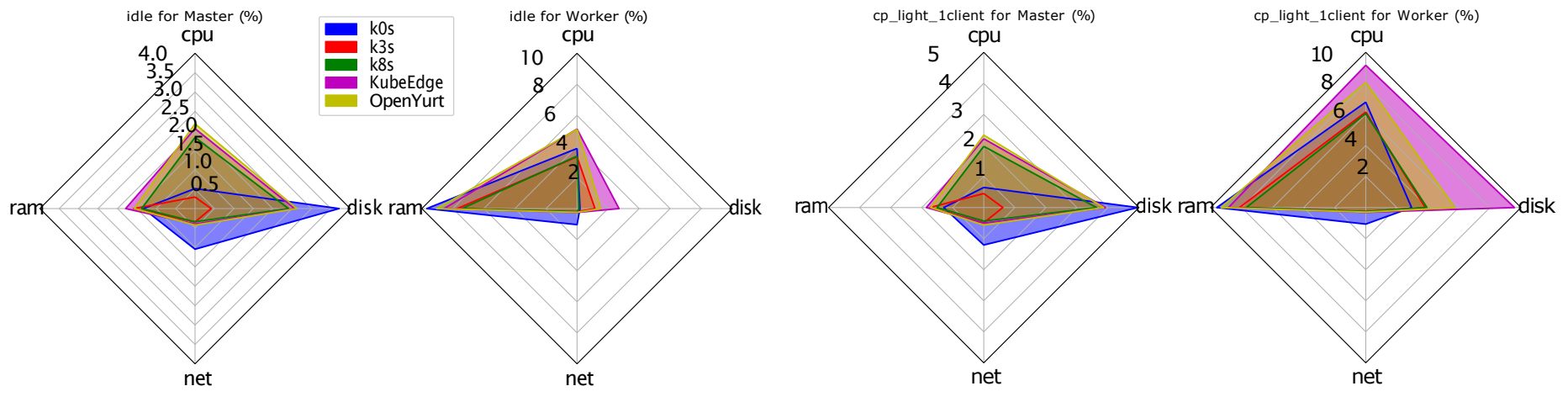

**Fig. 3.** Light tests: idle and cp_light_1client (Master and Worker nodes, %)

In conclusion, the resource efficiency of various lightweight KDs demonstrates notable differences under light load conditions. **(RQ1)** KubeEdge shows significantly higher CPU and Disk usage increases compared to k0s, k3s, k8s, and OpenYurt, indicating potentially lower processing efficiency and higher resource demands. **(RQ2)** Additionally, there are differences in scaling patterns that should be considered for optimized workload scheduling and resource management, particularly for resource-constrained environments.

## 4.2 Heavy tests

Adding additional workload significantly alters resource consumption. The most noticeable effects occur on worker nodes, with OpenYurt demonstrating the highest CPU consumption at 26%, followed by KubeEdge at 21%, and the less voracious k0s - 20%, k3s and k8s show 19% and 16% respectively. In terms of disk usage, KubeEdge is the most aggressive, showing Disk IO usage of 25%. Figure 4 presents a heavy test scenario for 12 clients. The results of testing another heavy cp_heavy_8client test could be found in the project's repository though the resource usage pattern is the same.

In the case of light tests, the time measurement is negligible since it doesn't make any differences between distributions. Nevertheless, the time dimension reveals a crucial distinction of kubernetes performances in heavy test scenarios.

10 https://kubernetes.io/docs/tasks/extend-kubernetes/setup-konnectivity/
11 https://docs.k0sproject.io/v1.21.0+k0s.0/networking/#controllers-worker-communication

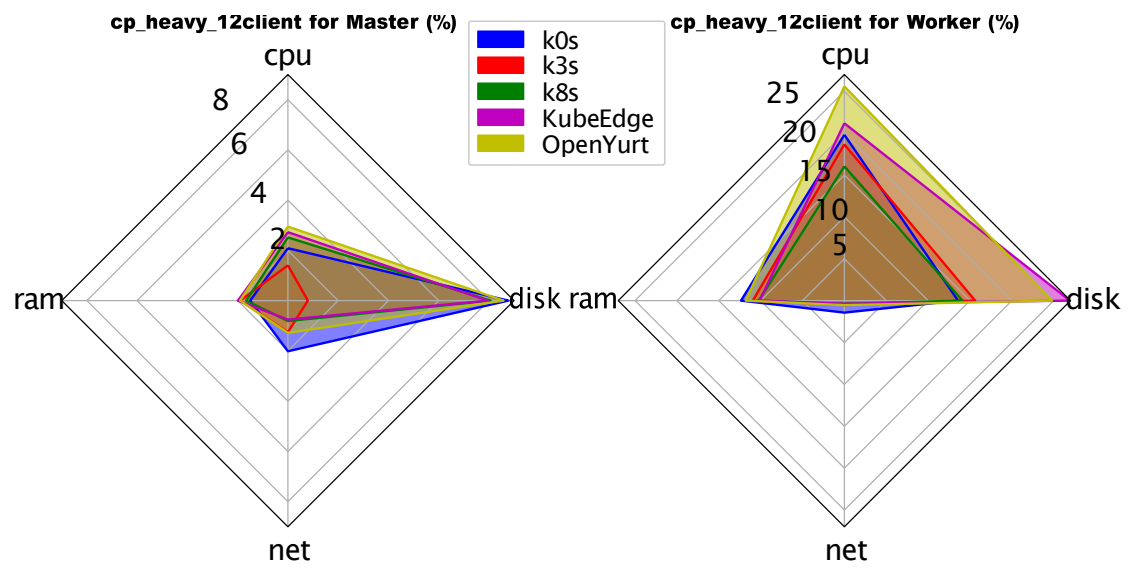

**Fig. 4.** Heavy tests: cp_heavy_12client (Master and Worker nodes, %)

Particularly, it shows the processing time of the same task could be accomplished by a specific KD, additionally, it shows the trajectory of a resource consumption. Moreover, the significance of kubernetes's time-based performance evaluation was not fully covered in [6], [12] works. Exemplifying diagram Figure 5 illustrates CPU usage of worker nodes under heavy load tasks executed on clusters exhibiting the differences in CPU utilization and more importantly the finish time of the workload. The iot-edge[12] project with various other diagrams could be viewed for more precise observations.

It is observable that KubeEdge takes a maximum of 10 min of processing time to accomplish heavy load tasks while OpenYurt and k8s do the same job within 6 min, and the fastest distributions become k0s and k3s with 4 min spent. While in heavy load tests, the most resource-efficient distribution is k8s, it is not the most performant distribution. The k0s and k3s do not demonstrate the most optimal resource usage, but it is clear that these lightweight distributions are more effective performance-wise, presumably due to less overhead compared to the rest.

To conclude, adding additional workload drastically alters resource consumption among the KDs. **(RQ1)** OpenYurt exhibits the highest CPU consumption under heavy loads, followed by KubeEdge, with k0s, k3s, and k8s showing lower CPU usage. **(RQ1)** KubeEdge also demonstrates the highest Disk IO usage, indicating its aggressive resource consumption under heavy load. **(RQ2)** Regarding performance, k0s and k3s complete heavy-load tasks faster than other distributions, despite not being the most resource-efficient. KubeEdge lags tremendously, 4 min from the nearest, indicating the slowest cluster under heavy load. **(RQ3)** These findings illustrate a trade-off between resource efficiency, performance and functionality, suggesting that the optimal choice of distribution depends on specific workload demands and resource availability.

12 https://github.com/DiyazY/iot-edge

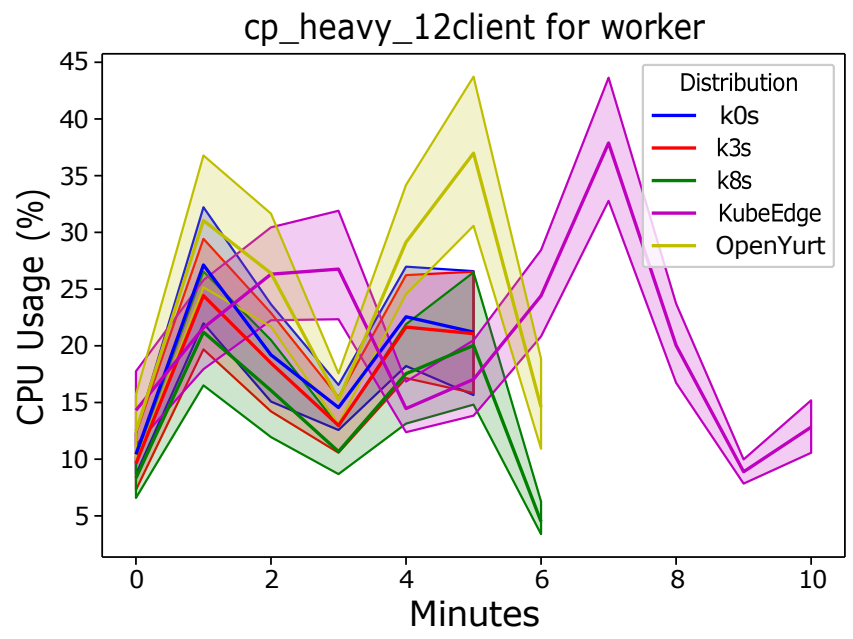

**Fig. 5.** CPU usage of worker nodes under cp_heavy_12client test

### 4.3 Operational metrics

Figure 6 illustrates the latency time for KDs scaling from 1 pod to 120 pods. All distributions scale in linear growth, where the pod increase results in pod startup delays gradually climbing from 1 sec to 7 sec respectively. However, the KubeEdge distributive does not scale as gracefully as others. Its latency for 10 pods already shows a significant gain, reaching almost 3 sec of delay. After, the latency rapidly jumps from 3 sec to 10 sec at 80 pods, which is a notable increase since it has already become the slowest system at 60 pods by overtaking OpenYurt distribution with its worst results at 120 pods. The upward trend is even more dramatic, the latency skyrockets to slightly above 30 sec at a point of scaling to 120 pods. If the scaling property is crucial, the KubeEdge is not a good alternative. Presumably, KubeEdge lags scaling by providing feature-rich services that put overheads.

Figure 7 depicts the latency and throughput for Pod Creation(PC) and Deployment (D). It shows relatively similar results for k0s, k3s and k8s, approximately 140 pods per min for PC workload and around 275 pods/min for D with latency between 2-4 ms. This is followed by OpenYurt showing 120 pods per minute for PC and 210 pods per minute for D with 3 ms and 8 ms of latency respectively. The slowest distribution in terms of PC throughput becomes KubeEdge since its maximum throughput is managed at around 60 pods per minute with the highest latency around 46 ms for D.

The throughput (pods/min) and latency (ms) of PC clearly show that the most performant distribution is k0s, which in general has the lowest latency and highest throughput, while for D, the most effective cluster is k8s. Just slightly behind those two is k3s, which in most cases show relatively similar results. Notably, the OpenYurt distribution is less efficacious compared to the previous 3. The worst results are shown by KubeEdge which is overall, 75% less effective (for D 50 pods/min vs 275 pods/min - k3s). It is always worth emphasizing that the latter two distributions are rich feature-wise, hence, it brings extra overhead on their shoulders, consequently, it may result in slower operations.

The performance analysis[13] of namespace, pod, deployment, and service operations reveals that k3s consistently demonstrate the lowest latency, making it optimal for performance-sensitive applications. In contrast, k0s, k8s, KubeEdge and OpenYurt exhibit higher latencies, suggesting potential inefficiencies likely due to additional overhead in managing operations. Particularly, get and list operations maintain low latencies across all environments, whereas create, update, and delete operations show greater variability, reflecting their complexity. These findings highlight k3s as a superior choice for environments demanding high performance, while k0s, k8s, KubeEdge and OpenYurt may be more suited for specialized use cases where their unique features justify the latency trade-offs.

[13] https://github.com/DiyazY/iot-edge/blob/main/src/diagrams/latency-statistics/cp_heavy_12client.pdf

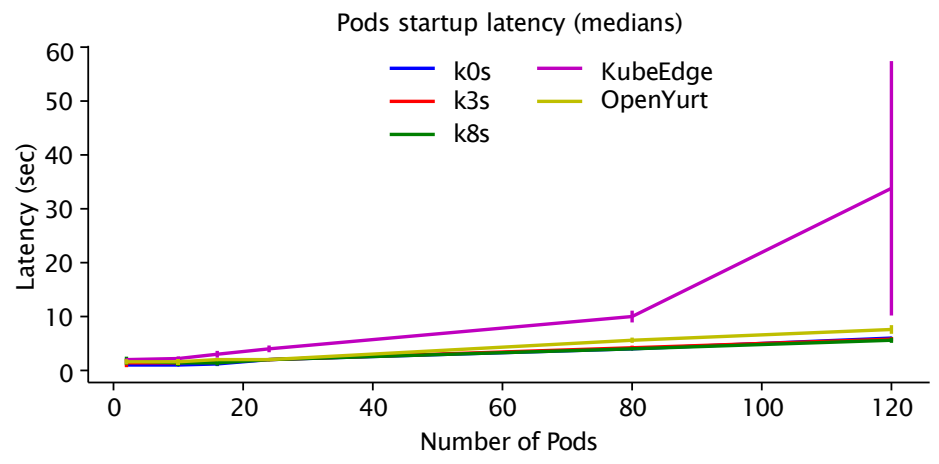

**Fig. 6.** Pod startup latency (medians)

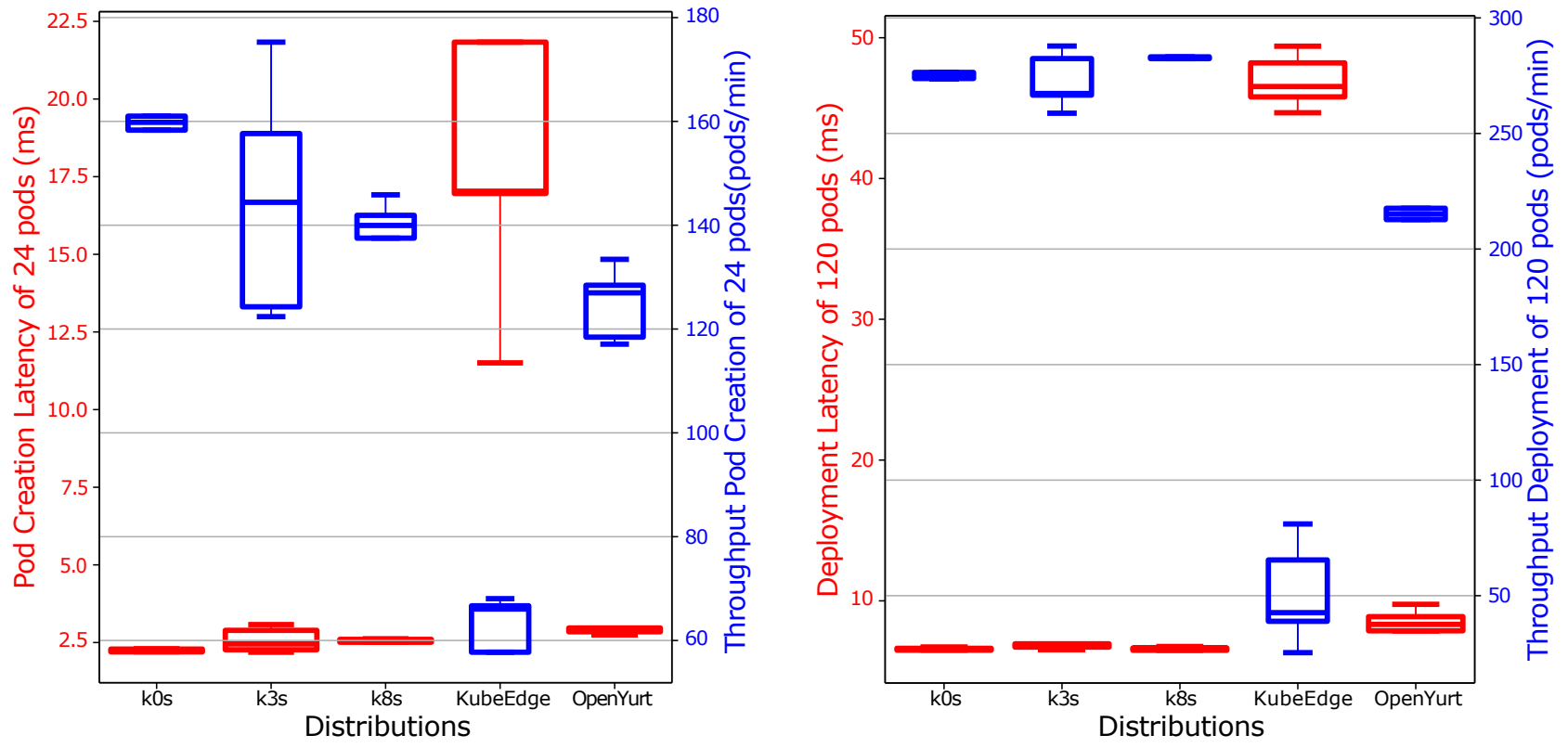

**Fig. 7.** cp_heavy_12client - Latency (ms) and Throughput (pods/min) for Pod Creation with 24 pods (left) and Deployment(right) with 120 pods (right).

Finally, operational metrics reveal significant differences in the scalability and performance of clusters. **(RQ2)** All distributions exhibit a linear increase in pod startup delays as they scale from 1 to 120 pods, with KubeEdge showing a notably higher latency increase, making it less suitable for scaling-intensive environments. **(RQ1)** In terms of pod creation throughput, k0s, k3s, and k8s demonstrate the highest performance, with k3s consistently showing the lowest latency for various operations. **(RQ3)** KubeEdge, while feature-rich, shows the worst performance with significantly higher latencies and lower throughputs, suggesting a trade-off between functionality and efficiency. **(RQ3)** These findings suggest that while k3s is optimal for operation performance-sensitive applications, k0s, k8s, KubeEdge, and OpenYurt may be more appropriate for specialized use cases where their unique features justify the latency trade-offs.

### 4.4 Data Plane

The data plane throughput, measured in operations per second (Ops/sec) using the memtier benchmark[14], is shown in Figure 8. This metric reflects the system's ability to handle data processing efficiently.

[14] https://github.com/RedisLabs/memtier_benchmark/releases/tag/2.0.0

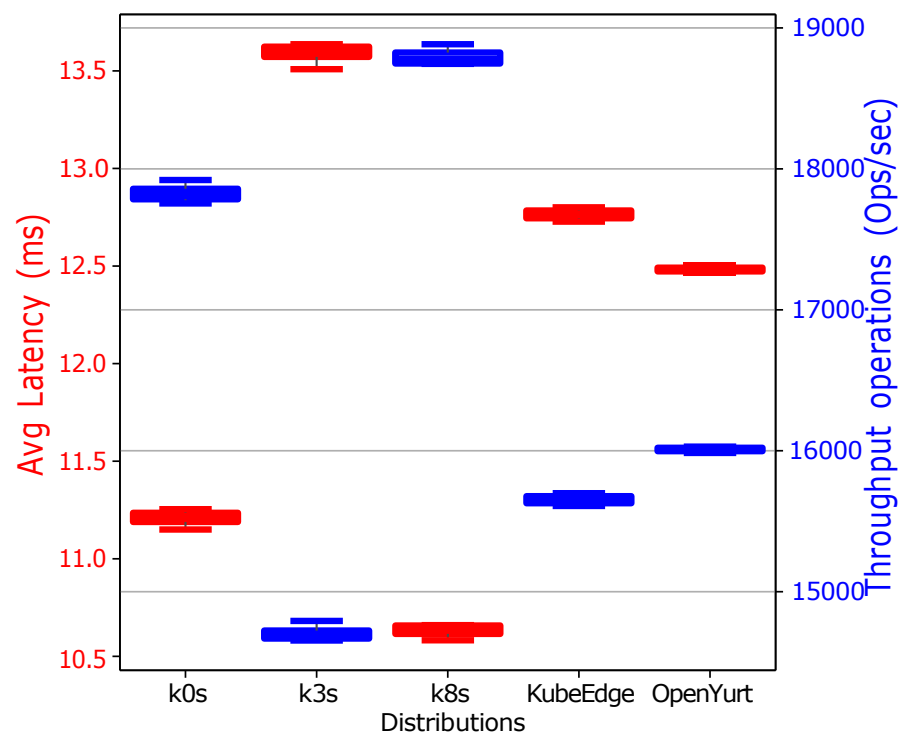

**Fig. 8.** Data plane - Average Latency (ms) and Throughput operations (Ops/sec)

k8s achieved the highest throughput, nearing 19,000 Ops/sec, followed by k0s at around 18,000 Ops/sec. OpenYurt reached just over 16,000 Ops/sec, while KubeEdge slightly trailed at under 16,000 Ops/sec. k3s had the lowest throughput, below 15,000 Ops/sec.

Latency, depicted in Figure 8, measures the average response time. k8s had the lowest latency at 10.5 ms, followed by k0s at just above 11.0 ms. OpenYurt and KubeEdge were similar, with latencies slightly under and over 12.5 ms, respectively. k3s recorded the highest latency, around 13.5 ms.

When comparing both throughput and latency, k8s consistently outperforms other distributions, making it the most efficient option for data-heavy high-performance requirements. It achieves the highest throughput and the lowest latency, signifying a robust and responsive data plane. Conversely, k3s shows the least favorable results in both metrics, suggesting potential areas for optimization in data plane throughput and latency.

k0s, while not as performant as k8s, still offers commendable efficiency with high throughput and relatively low latency. OpenYurt maintains a balanced performance but does not excel in either metric compared to the more optimized distributions. KubeEdge, designed for edge computing, shows reduced performance in both throughput and latency, which might be a trade-off for other edge-specific benefits not captured in these metrics.

Resource utilization-wise, clusters' behaviors tangibly differentiate on master machines while remaining almost the same across worker machines (Figure 9). For master machines, k0s, k3s and OpenYurt show lower CPU usage (0.5% - 1.1%) compared to k8s, KubeEdge (1.7%-2%), indicating better efficiency. The RAM consumption, k3s uses the least RAM, followed by k8s, OpenYurt, k0s and KubeEdge, which uses the most. Network utilization is relatively low across all distributions, except k3s being notably more network-heavy. Disk usage adheres to a similar pattern, it is almost the same for k0s, k8s, OpenYurt and KubeEdge, around 2.5%, and significantly higher for k3s by showing 3.5% consumption.

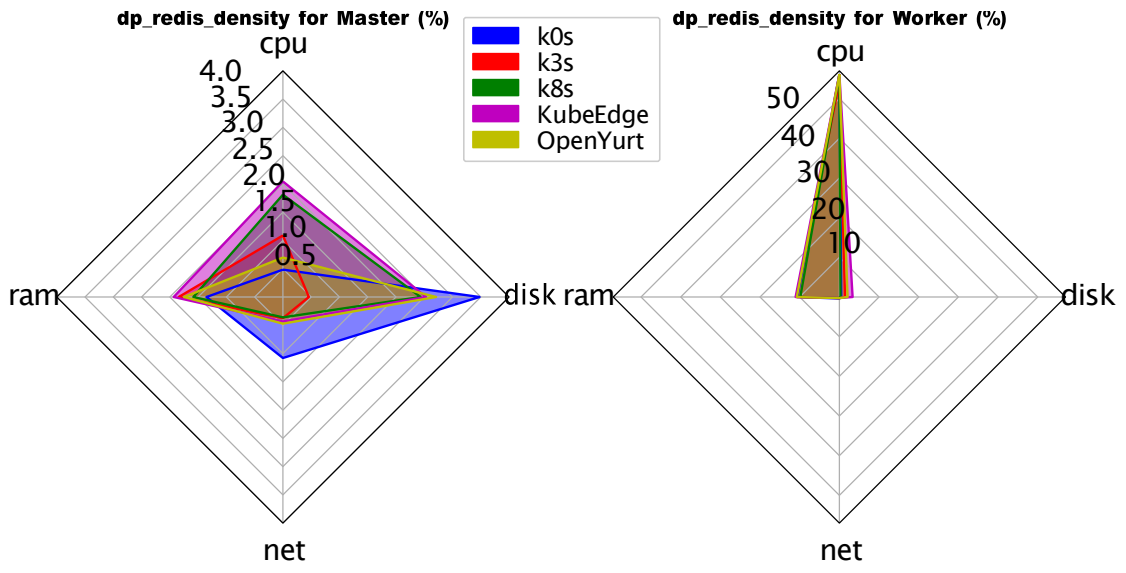

**Fig. 9.** dp_redis_density - performance metrics (%)

In summary, k8s is the best for high-performance needs **(RQ2)**, while k0s is efficient for resource-constrained environments. k3s and KubeEdge require optimization, and OpenYurt is balanced but not ideal for peak performance. **(RQ3)** These findings emphasize the need to align performance requirements and resource efficiency when choosing the appropriate KD.

## 5 Discussion

This analysis of KDs reveals a spectrum of resource usage and performance trade-offs. k0s and k8s provide the best balance of throughput and resource efficiency, ideal for demanding applications. k3s, while efficient, requires enhancements for high-throughput environments. KubeEdge and OpenYurt, although feature-rich for edge computing, exhibit higher resource demands, underscoring the need for further refinement to improve their performance and efficiency. **(RQ3)** Table 3 summarizes our findings.

### 5.1 Future Challenges

While our study provides valuable insights into the performance and resource efficiency of lightweight KDs, several challenges remain to optimize their deployment in edge computing environments. Future work should focus on:

- Optimizing Edge-Specific Distributions: Enhancing the performance and reducing the resource consumption of KubeEdge and OpenYurt without compromising their edge-specific functionalities.
- Scalability Improvements: Investigating methods to improve the scalability of KDs, particularly for KubeEdge, to better handle increased workloads and larger numbers of pods.
- Integration of IoT Features with Efficiency: Developing strategies to integrate IoT-specific features into lightweight KDs like k3s and k0s while maintaining low resource consumption.
- Standardized Benchmarking Tools: Establishing standardized tools and methodologies for benchmarking KDs in edge computing environments to facilitate more consistent and comparable results across studies.

**Table 3.** Summary Comparison of Kubernetes Distributions

| Dist. | Strengths | Weaknesses | Resource Consumption | Suggestion |
|---|---|---|---|---|
| k0s | High throughput and low latency with minimal resource consumption. Simple installation and management. | Fewer features compared to full Kuernetes distributions. Limited community and enterprise support. | Moderate: efficient in disk usage; moderate CPU and RAM usage. | Highly efficient for resource-constrained environments needing balanced performance and simplicity. Ideal for edge and IoT deployments. |
| k3s | Extremely lightweight with a small binary size. Low memory and CPU footprint. Simple installation and management. | Lower throughput and higher latency for data-heavy conditions. Limited advanced features and extensions. | Low: lowest utilization on Master node and moderate on Worker nodes. | Optimal for very resource-constrained environments with some compromises in performance under data-heavy load. |
| k8s | Highest throughput and lowest latency. Robust ecosystem with extensive tools and features. Highly scalable and flexible. | Complex to set up and manage. Overhead can be significant for small-scale deployments. | Low: moderate usage on Master node and low usage on Worker nodes. | Best suited for high-performance, large-scale environments. |
| OpenYurt | Seamless integration between cloud and edge environments. Enhanced capabilities for managing distributed IoT applications. | Moderate throughput and higher latency. Additional overhead for edge-specific features. Complex to set up and manage. | High: high disk and RAM consumption with the highest CPU utilization. | Well-suited for hybrid cloud-edge environments with IoT applications, balancing performance with advanced features. |
| KubeEdge | Tailored for edge computing with offline autonomy. Supports IoT device management and communication. | Higher resource consumption compared to other lightweight distributions. Lower throughput and higher latency. Complex to set up and manage. | High: high CPU and RAM utilization with highest disk utilization on Worker nodes. | Best for specialized edge computing scenarios where offline capabilities and IoT integrations are critical, though resource optimization is needed. |

## 6 Conclusion

This paper analyzed the performance and resource utilization of lightweight KDs, focusing on k0s, k3s, k8s, OpenYurt, and KubeEdge. Our findings show that k8s and k0s are the most efficient, balancing performance and resource usage, particularly under both light and heavy loads. KubeEdge, while tailored for edge computing, consumes more CPU and disk resources, making it less ideal for resource-constrained environments. k3s, though resource-efficient, requires optimization for better throughput and latency in data-heavy scenarios.

Selecting the right KD depends on specific deployment needs. k8s and k0s are recommended for high-performance environments, while k3s, with further tuning, could be effective in resource-limited settings. KubeEdge and OpenYurt, although feature-rich, may need additional optimization.

Future research should focus on the unique optimizations of each distribution, particularly those designed for edge computing, to further enhance their efficiency and applicability. By understanding these factors, kubernetes deployments can be better optimized for diverse environments, ensuring efficient resource use and high performance.

## Acknowledgement

This work is supported by funding from Business Finland in project Industry X.